\documentclass[twocolumn,showpacs,preprintnumbers,nofootinbib,amsmath,amssymb,aps,prd,superscriptaddress]{revtex4}
\usepackage{amsmath,amssymb}
\usepackage{amsfonts,amssymb,mathrsfs}
\usepackage[bookmarks=false,pdfstartview=FitH]{hyperref} 

\def\be{\begin{equation}}
\def\ee{\end{equation}}
\def\nn{\nonumber}
\def\f{\frac}
\def\tf{\tfrac}

\def\pl{{\rm Pl}}
\def\lp{\ell_\pl}
\def\b{\bar}
\def\d{\dot}

\def\v{\vec}

\def\dd{{\rm d}}
\def\na{\nabla}

\def\de{\delta}
\def\ep{\epsilon}

\def\om{\omega}

\def\vp{\varphi}
\def\De{\Delta}
\def\mR{\mathcal{R}}

\begin{document}

\pagestyle{plain}

\title{The Matter Bounce Scenario in Loop Quantum Cosmology}

\author{Edward Wilson-Ewing} \email{wilson-ewing@cpt.univ-mrs.fr}

\affiliation{Aix-Marseille Universit\'e, CNRS UMR 7332, CPT, 13288 Marseille, France}
\affiliation{Universit\'e de Toulon, CNRS UMR 7332, CPT, 83957 La Garde, France}

\begin{abstract}

In the matter bounce scenario, a dust-dominated
contracting space-time generates scale-invariant
perturbations that, assuming a nonsingular
bouncing cosmology, propagate to the expanding
branch and set appropriate initial conditions
for the radiation-dominated era.  Since this
scenario depends on the presence of a bounce,
it seems appropriate to consider it in the
context of loop quantum cosmology where a bouncing
universe naturally arises.  For a pressureless
collapsing universe in loop quantum cosmology, the
predicted power spectrum of the scalar perturbations
after the bounce is scale-invariant and the tensor
to scalar ratio is negligibly small.  A slight red
tilt can be given to the scale-invariance of
the scalar perturbations by a scalar field whose
equation of state is $P = - \epsilon \rho$, where
$\epsilon$ is a small positive number.  Then, the
power spectrum for tensor perturbations is also
almost scale-invariant with the same red tilt as
the scalar perturbations, and the tensor to scalar
ratio is expected to be $r \approx 9 \times 10^{-4}$.
Finally, for the predicted amplitude of the scalar
perturbations to agree with observations, the critical
density in loop quantum cosmology must be of the order
$\rho_c \sim 10^{-9} \rho_{\rm Pl}$.

\end{abstract}


\maketitle

\section{Introduction}
\label{s.intro}

Observations of the cosmic microwave background (CMB) have shown that the
spectrum of scalar perturbations is scale-invariant with a slight red tilt
\cite{Komatsu:2010fb} and therefore this is one of the major predictions
that any viable cosmological model must make.

There are two well known mechanisms that generate scale invariant
perturbations starting from the fluctuations of an initial quantum vacuum
state: inflation, an exponential expansion of the universe, and
a dust-dominated contracting phase.  Recent reviews of these two
paradigms are \cite{Baumann:2009ds} and \cite{Brandenberger:2012zb},
respectively.  The fact that they both give rise to a scale-invariant
spectra of scalar and tensor perturbations is easily understood as they
are related by a simple duality \cite{Wands:1998yp}.

In this paper, we will focus on the second mechanism.  More specifically,
we will study the matter bounce scenario where an initially classical
contracting dust-dominated universe with quantum vacuum fluctuations
gives scale-invariant perturbations once the relevant modes exit the
Hubble radius.  Assuming the presence of a bounce, the scale-invariant
perturbations then provide appropriate initial conditions for the
expanding radiation-dominated epoch of our universe.

However, the standard classical treatment of the matter bounce has a
major shortcoming in that one must evolve the perturbations from the
pre-bounce era to the post-bounce one by hand.  Typically this is done
by using some reasonable matching conditions \cite{Finelli:2001sr},
but it would be nice to go beyond this and explicitly calculate the
propagation of the perturbations through the bounce, where quantum
gravity effects are expected to be important.  Now, since the matter
bounce scenario depends on the presence of a bounce, it is natural to
work in the context of loop quantum cosmology (LQC) which predicts a
bouncing universe.  In this paper, we will study the matter bounce in
LQC in order to determine the consequences of quantum gravity effects
in this setting.  As we shall see, some of the results obtained here
are significantly different from those obtained in the matching
procedure of \cite{Finelli:2001sr}; these differences are due to
modifications of the Friedmann and Mukhanov-Sasaki equations from
quantum gravity effects.

Homogeneous LQC is obtained by following loop quantum gravity (LQG) and
using holonomy and flux operators in order to quantize the Hamiltonian
constraint corresponding to a homogeneous and isotropic space-time.
See \cite{Ashtekar:2011ni, Banerjee:2011qu} for recent reviews of LQC.
One of the main results of LQC is that the classical big-bang singularity
is resolved and replaced by a bounce that occurs when the space-time
curvature is approximately of the Planck scale.  In addition, careful
studies of the Friedmann-Lema\^itre-Robertson-Walker (FLRW) cosmologies
in LQC have shown that, for states that are semi-classical (i.e.,
sharply peaked around a classical solution at late times), there exist
a set of effective equations that provide an excellent approximation
to the full quantum dynamics of the states at all times, even during
the bounce \cite{Ashtekar:2006wn, Taveras:2008ke}.

Cosmological perturbations have also been studied in some detail in LQC,
both at the quantum level \cite{WilsonEwing:2012bx} and especially in the
effective theory \cite{WilsonEwing:2011es, Cailleteau:2011kr, Cailleteau:2011mi,
Cailleteau:2012fy}.  In this paper we will work with the effective equations
for the sake of simplicity.  This clearly requires the assumption that the
effective equations continue to be a good approximation to the quantum dynamics
of semi-classical states in the presence of linear perturbations, which seems
reasonable so long as the perturbations remain small.

As an aside, we point out that there are two types of corrections, holonomy
and inverse triad, that are considered in effective studies in LQC.  In
this paper we focus on holonomy corrections as they are the dominant ones
in the quantum theory of homogeneous LQC and we assume this will continue
to be the case when linear perturbations are included.  See however
\cite{Bojowald:2011iq} for a different perspective.

The paper is organized as follows: in Sec.\ \ref{s.lqc} we review the effective
theory in LQC for both the homogeneous FLRW background and the perturbations.
Then we study the dynamics of the cosmological perturbations in a bouncing
dust-dominated FLRW universe; scalar perturbations in Sec.\ \ref{s.scal}
and tensor perturbations in Sec.\ \ref{s.tens}.  We close with a discussion
in Sec.\ \ref{s.dis}.

\section{Loop Quantum Cosmology}
\label{s.lqc}

We start by briefly introducing the ingredients coming from loop quantum
cosmology that will be necessary for the calculations in the following
sections, namely the LQC effective equations.  The effective equations
are derived from the LQC Hamiltonian constraint operator and include
the leading order quantum gravity corrections to the classical Friedmann
and Mukhanov-Sasaki equations coming from LQC.  It turns out that the
effective equations provide a surprisingly good approximation to the
dynamics of sharply peaked states in LQC at all times, including at
the bounce point where quantum gravity effects are strongest.  

Since we are only interested in space-times that are classical (and
therefore sharply peaked) at times well before and well after the bounce,
we can restrict our attention to semi-classical states in LQC.  Because
of this we can work in the effective theory, which will considerably
simplify the analysis.

An early derivation of the effective equations for the flat FLRW space-time
is given in Appendix B of \cite{Ashtekar:2006wn}, while a considerably
more detailed study can be found in \cite{Taveras:2008ke}.  Recent reviews
of effective equations, including those for some other cosmological
space-times, are in Sec.\ V of \cite{Ashtekar:2011ni} and Part III
of \cite{Banerjee:2011qu}.

We will review the effective equations for the flat FLRW background in the
first part of this section, and the effective theory for the scalar and
tensor perturbations in the second part.

\subsection{The Homogeneous Background}
\label{ss.lqc-hom}

Given the flat FLRW metric in terms of the proper time,
\be 
\dd s^2 = - \dd t^2 + a(t)^2 \dd \v{x}^2,
\ee
where $a(t)$ is the scale factor, and denoting the energy density of
the matter field by $\rho$ and its pressure by $P$, the LQC effective
equations for the background, including corrections due to quantum
geometry effects, are
\begin{align} \label{lqc-fried}
H^2 &= \f{8 \pi G}{3} \rho \left( 1 - \f{\rho}{\rho_c} \right), \\
\f{\ddot{a}}{a} - H^2 &= - 4 \pi G \left( \rho + P \right)
\left( 1 - \f{2 \rho}{\rho_c} \right), \\
\d{\rho} &= - 3 H \left( \rho + P \right),
\end{align}
where $\rho_c$ is the critical energy density and $H = \d{a}/a$ is
the Hubble rate.  The dot denotes differentiation with respect to
the proper time $t$.  Finally, note that the classical Friedmann
equations are recovered in the limit of $\rho_c \to \infty$.

When the matter field is a perfect fluid with a constant equation of
state $P = \om \rho$, these equations can be solved, giving
\begin{align} 
\rho &= \rho_o a^{-3(1+\om)}, \\
\label{lqc-a}
a(t) &= \left( 6 \pi G \rho_o (1+\om)^2 (t - t_o)^2
+ \f{\rho_o}{\rho_c} \right)^\f{1}{3(1+\om)}.
\end{align}
From the nonvanishing form of $a(t)$, it is clear that the big bang
singularity is resolved and replaced by a bounce.  The dimensionful
constants of integration $\rho_o$ and $t_o$, which determine the
magnitude of the scale factor and the time the bounce occurs at,
will be set to $\rho_o = \rho_c$ and $t_o = 0$ so that
$a(t_{\rm bounce}) = 1$ and $t_{\rm bounce} = 0$; these choices
do not affect the physics.

As an aside, it is worth pointing out that while the scale factor $a$
can be expressed in a relatively simple manner in terms of the proper
time $t$, this is not true when one works in conformal time $\eta$,
defined by
\be \label{dt-deta}
\dd \eta = \f{\dd t}{a(t)},
\ee
in which case the form of the scale factor is considerably more complicated.
For this reason, we will try to work in proper time whenever LQC effects
are important.  On the other hand, it will be useful to work in conformal
time in order to study the perturbations in the classical limit and
therefore we will switch between the two different time choices depending
on the situation.

Finally, if one is working with a scalar field, then the energy density and
the pressure are given by the same relations as in the classical theory,
\be 
\rho = \f{1}{2} \, \d\vp^2 + V(\vp), \qquad
P = \f{1}{2} \, \d\vp^2 - V(\vp).
\ee
It is possible, for the potential%
\footnote{Note that the denominator in the potential
is necessary due to the modifications in the LQC
Friedmann equations.  The usual potential used
to get a constant equation of state given in e.g.\
\cite{Finelli:2001sr} is obtained in the limit
$\rho_c \to \infty$.}
\be \label{potential}
V(\vp) = \f{(1 - \om) V_o e^{-\sqrt{24 \pi G (1+\om)} \vp}}
{\left( 1 + \tf{V_o}{2 \rho_c} e^{-\sqrt{24 \pi G (1+\om)} \vp} \right)^2},
\ee
to mimic a cosmology with a constant equation of state $P = \om \rho$
by using a scalar field, so long as one starts with appropriate
initial conditions \cite{Mielczarek:2008qw}.  This is what will be
done here, for the case of $\om = 0$.

\subsection{Perturbations}
\label{ss.lqc-perts}

Classically, scalar perturbations in cosmology can be studied by using the
Mukhanov-Sasaki equation.  (For an introduction to the theory of cosmological
perturbations in the classical theory, see for example \cite{Mukhanov:1990me,
Peter-Uzan}.)  In LQC, there are modifications to the equations of motion
governing the dynamics of the perturbations, just as the Friedmann
equations are modified.  The corrections have been studied for the cases
of a perfect fluid \cite{WilsonEwing:2011es} and a scalar field
\cite{Cailleteau:2011kr, Cailleteau:2011mi}.  Here we are working with
a scalar field, in which case the modified Mukhanov-Sasaki equation is
\be 
v'' - \left( 1 - \f{2 \rho}{\rho_c} \right) \na^2 v
- \f{z''}{z} v = 0,
\ee
where the prime denotes differentiation with respect to the conformal time
$\eta$,
\be 
v = a \, \sqrt{\rho + P} \, \de u^{(gi)} - z \Phi,
\ee
where $\de u^{(gi)}$ and $\Phi$ are the usual gauge-invariant observables in
cosmological perturbation theory and
\be \label{def-z}
z = \f{a}{H} \sqrt{ \left( \rho + P \right) }.
\ee
Note that although \eqref{def-z} has the same form as in classical general
relativity, the dynamics of all of the background variables are modified in
LQC and $z$ will behave very differently near the bounce as compared to what
might be expected classically.  In particular, if the background dynamics
are those of a perfect fluid with a constant equation of state $P = \om \rho$
as is the case for the matter bounce scenario, then Eq.\ \eqref{lqc-fried}
shows that there is an extra term coming from LQC in the denominator,
\be \label{scal-z}
z = \sqrt{ \f{3 (1 + \om)}{8 \pi G}} \cdot \f{a}{\sqrt{1 - \rho / \rho_c}}.
\ee

The Mukhanov-Sasaki equation is most easily solved in Fourier space, where it
becomes
\be \label{lqc-ms}
v_k'' + \left( 1 - \f{2 \rho}{\rho_c} \right) k^2 v_k - \f{z''}{z} v_k = 0.
\ee

Tensor perturbations behave a little differently: their dynamics are governed
by an equation that has the same form as for scalar perturbations, but even
classically the variables are defined slightly differently, and it turns out
that the quantum geometry corrections do not appear in exactly the same
manner either.  The holonomy-corrected Mukhanov-Sasaki equation for tensor
perturbations in Fourier space is given by \cite{Cailleteau:2012fy}
\be \label{lqc-tens}
\mu_k'' + \left( 1 - \f{2 \rho}{\rho_c} \right) k^2 \mu_k - \f{z_T''}{z_T} \mu_k = 0,
\ee
where
\be \label{zt}
z_T = \f{a}{\sqrt{1 - 2 \rho / \rho_c}},
\ee
and $\mu_k = z_T h_k$, with $h$ being the usual gravitational wave
perturbation variable.  Note that there is no need for absolute values in
the denominator of $z_T$ as $\mu_k$ is a complex variable.  The fact that
there are no absolute values will be important in Sec.\ \ref{s.tens}.

Finally, one might be worried about the divergences in $z$ and $z_T$ that
occur at or near the bounce point.  However, we shall see that the
solutions for $v_k$ and $\mu_k$ can be determined despite these
divergences and, assuming small perturbations before the bounce, the
perturbations are small after the bounce as well.  Therefore, these
divergences do not prevent us from obtaining explicit solutions for $v_k$
and $\mu_k$ and studying their properties after the bounce.

A separate open question is whether the divergences in $z$ and $z_T$ can
drive the perturbations to become large for a short period of time close
to the bounce, at which point back reaction effects might become important.
We leave a more careful study of this issue for future work.

For ease of notation, we will drop the index $k$ in \eqref{lqc-ms} and
\eqref{lqc-tens} in the future.

\section{Scalar Perturbations}
\label{s.scal}

We will begin by studying the propagation of scalar perturbations in an
LQC dust-dominated universe.
As perturbations are most easily studied in conformal time, and the initial
conditions are to be set in a classical regime where quantum gravity effects
are negligible, it is useful to recall the classical relations giving
the scale factor and the proper time in terms of the conformal time for
$P = 0$,
\be \label{a-eta}
a(\eta) = \f{2 \pi G \rho_c}{3} \, \eta^2, \qquad
t(\eta) = \f{2 \pi G \rho_c}{9} \, \eta^3.
\ee
Another useful relation is $z(\eta)$ in the classical limit, given by
\be 
z(\eta) = \sqrt \f{\pi G}{6} \, \rho_c \, \eta^2.
\ee

We will now study how the perturbations propagate through the bounce
by imposing quantum vacuum initial conditions in the pre-bounce phase
and then determining the form of the perturbations in the post-bounce
phase.  This will be done by matching the classical solutions, in
both the pre- and post-bounce eras, order by order with a formal
solution given by an expansion in $k$ that holds at all times,
including at the bounce point.

\subsection{The Contracting Branch and the Bounce}
\label{ss.scal-cont}

In the classical limit, and for a dust-dominated universe, \eqref{lqc-ms}
simplifies to
\be \label{diff-bessel}
v'' + \left( k^2 - \f{2}{\eta^2} \right) v = 0.
\ee
If \eqref{diff-bessel} is rewritten for $f = v / \sqrt{-\eta}$
and the time variable is rescaled by a factor of $k$, this becomes
the Bessel differential equation for $f$, and thus the solution to
\eqref{diff-bessel} is given by
\be \label{sol-bessel}
v(\eta) = \sqrt{-\eta} \left[ A_1 H_{3/2}^{(1)} (-k \eta) +
A_2 H_{3/2}^{(2)} (-k \eta) \right],
\ee
where $H_n^{(1)}(x)$ and $H_n^{(2)}(x)$ are the Hankel functions,
while $A_1$ and $A_2$ are constants to be determined by the initial
conditions.  Note the presence of a minus sign in front of $\eta$
as this solution holds in the contracting branch, where the time
variables are negative.

Now, by the asymptotic behaviour of the Hankel functions, we find
that (up to an irrelevant global phase which we leave out)
\be 
\lim_{\eta \to -\infty} v(\eta) \sim A_1 \sqrt{ \f{2}{\pi k} } \, e^{-i k \eta}
+ A_2 \sqrt{ \f{2}{\pi k} } \, e^{i k \eta},
\ee
and therefore it is possible to impose the vacuum initial conditions
\be \label{initial}
v_{\rm initial} = \sqrt \f{\hbar}{2 k} e^{-i k \eta},
\ee
by choosing $A_1 = \sqrt{\pi \hbar / 4}$ and $A_2 = 0$, which gives
\be \label{v-initial}
v(\eta) = \sqrt{ \f{-\pi \hbar \eta}{4} } H_{3/2}^{(1)} (-k \eta).
\ee
This gives the expression of the Mukhanov variable so long as
quantum geometry effects are negligible.

Of course, as the space-time contracts, the curvature will increase
and at some point LQC effects will begin to play an important role.
Therefore it will be necessary to match this solution to another
one which will be valid throughout the bounce.

This can be done by rewriting \eqref{lqc-ms} in an integral form,
i.e., as
\begin{align} \label{v-formal}
v(\eta) =& B_1 z + B_2 z \int^\eta \f{\dd \b\eta}{z^2}
- k^2 z \int^\eta \f{\dd \b\eta}{z^2}
\int^{\b\eta} \dd \b{\b\eta} \, z \, v \nn \\ &
+ \f{2 k^2}{\rho_c} z \int^\eta \f{\dd \b\eta}{z^2}
\int^{\b\eta} \dd \b{\b\eta} \, \rho \, z \, v.
\end{align}
Doing an expansion in $k$, we immediately find that the leading
order terms are
\be 
v(\eta) = B_1 \left( z + O(k^2) \right)
+ B_2 \left( z \int^\eta \f{\dd \b\eta}{z^2} + O(k^2) \right).
\ee
An important point here is that the constants $B_1$ and $B_2$
can (and will) depend on $k$.  For a dust-dominated space-time
in LQC,
\be \label{rad-az}
a(t) = \left( 6 \pi G \rho_c t^2 + 1 \right)^{1/3}, \quad
z(t) = \f{a(t)^{5/2}}{4 \pi G \sqrt{\rho_c} \, t},
\ee
and by using \eqref{dt-deta} we find that the two leading order terms
in $k$ of $v(t)$ are given by
\begin{align} \label{v-bounce}
v(t) = \, & B_1 z(t)
+ \Bigg[ \sqrt \f{8 \pi G}{27 \rho_c} \left(
\arctan \sqrt{6 \pi G \rho_c} t + \f{\pi}{2} \right) \nn \\ & \qquad
- \f{4 \pi G t}{3 \left( 6 \pi G \rho_c t^2 + 1 \right)} \Bigg] B_2 z(t),
\end{align}
where the constant of integration has been chosen in order to
simplify the matching of this solution with \eqref{v-initial}.
Indeed, in the classical regime of the contracting branch where
$t \ll -1 / \sqrt{6 \pi G \rho_c}$, we find that
\be 
v(t) = B_1 z(t) - \f{4 B_2 z(t)}{9 \rho_c t},
\ee
and it is possible to match this solution with \eqref{v-initial}
by ensuring that the coefficients of the leading terms in $k$
match in the classical regime.

We cannot take the limit $\eta \to 0$ of \eqref{v-initial} as
quantum gravity effects are important in this regime and then
the classical solution cannot be trusted.  Instead, we will
consider the limit of small $k \eta$ at a time where general
relativity can be trusted.  Note that this limit can only be
taken for the modes that have become larger than the Hubble
radius before LQC effects become important.  However, in the
matter bounce scenario all of the modes that we can observe
in the CMB today satisfy this property and therefore this
limitation is not problematic.  In the small $k \eta$ limit,
$v(\eta)$ contributes two terms, each of which can be used
to fix one of $B_1$ and $B_2$.

In the $k \eta \ll 1$ limit, the solution \eqref{v-initial}
tends to
\be 
v(\eta) \to \f{\sqrt{\hbar}}{3 \sqrt{2}} \, k^{3/2} \eta^2
- \f{i\sqrt{\hbar}}{\sqrt{2} \, k^{3/2} \, \eta},
\ee
and therefore the constants $B_1$ and $B_2$ must be taken to be
\be 
B_1 = \f{\sqrt{\hbar}}{\sqrt{3 \pi G} \rho_c} \, k^{3/2}, \quad
B_2 = i \f{\sqrt{3 \pi G \hbar}}{2} \, \rho_c \, k^{-3/2}.
\ee

\subsection{The Expanding Branch}
\label{ss.scal-exp}

We now know the form of $v(t)$, at least in the form of
the formal solution \eqref{v-formal}, and from this it is
possible to determine the spectrum of scalar perturbations
after the bounce.

In order to do this, we will again work with the leading
order terms of the formal solution to $v$, given by
\eqref{v-bounce}.  For times after the bounce where
quantum gravity effects are negligible,
$t \gg 1 / \sqrt{6 \pi G \rho_c}$ and then the behaviour
of $v(t)$ is, for the leading order terms in $k$,
\be \label{v-postB}
v(t) = \left( B_1 + \sqrt \f{8 \pi^3 G}{27 \rho_c} B_2 \right) z(t)
- \f{4 B_2 z(t)}{9 \rho_c t},
\ee
and we see that a mixing has occurred in the first term.
It is this new term, coming from the mixing, that will give
a scale-invariant spectrum.  This is what we shall show now.

In order to have the full solution to $v(\eta)$, the easiest
method is to use the classical Mukhanov-Sasaki equation
\eqref{diff-bessel} that holds once LQC effects become
small after the bounce.  The classical solution, to all
orders in $k$, is given by
\be 
v(\eta) = \sqrt{\eta} \left[ C_1 H_{3/2}^{(1)} (k \eta)
+ C_2 H_{3/2}^{(2)} (k \eta) \right],
\ee
where $C_1$ and $C_2$ are two constants that must be
determined by matching the lowest order terms in $k$ of
this solution with \eqref{v-postB}.

This matching gives
\be \label{v-final}
v(\eta) = \sqrt{\pi \hbar \eta} \left[ \f{1}{2} H_{3/2}^{(2)} (k \eta) 
+ i \f{\pi^2 (G \rho_c)^{3/2}\!\!\!}{\sqrt{6} \: k^3} J_{3/2}(k \eta) \right],
\ee
where $J_n(x)$ is the Bessel function of the first kind and it
is understood that the prefactors have been determined to leading
order in $k$.

Therefore, at times after the bounce when quantum gravity effects
are negligible (but before the modes reenter the Hubble
radius), the perturbations look like
\begin{align}
v(\eta) = \: & \f{i \, \sqrt{\hbar}}{k^{3/2}} \left[
\pi^{5/2} \left( \f{G \rho_c}{3} \right)^{3/2} \!\! \eta^2
- \f{1}{\sqrt{2} \pi \eta} \right] \nn \\ &
+ \f{\sqrt{\hbar}}{3 \sqrt{2} \pi} k^{3/2} \eta^2.
\end{align}
In order to calculate the amplitude of the scalar fluctuations, we can use
the variable $\mR$ which measures the curvature perturbations and is related
to $v$ simply by \cite{Mukhanov:1990me, Peter-Uzan},
\be \label{mR-final}
\mR(\eta) = \f{v}{z} \sim i \f{\sqrt{2 \pi^4 G^2 \hbar \rho_c}}{3} k^{-3/2},
\ee
where we have only kept the dominant mode in terms of $k$, and also dropped
the mode that decays as $\eta$ grows.  From this, it is possible to calculate
the scalar power spectrum given by
\be \label{amp-scal}
\De^2_\mR(k) = \f{k^3}{2 \pi^2} |\mR(\eta)|^2 \sim \f{\pi^2 G^2 \hbar \rho_c}{9} \, k^0.
\ee
This shows that the scalar spectral index is $n_s = 1$, and thus the
power spectrum is scale-invariant.

For the amplitude of the scalar perturbations given in \eqref{amp-scal}
to agree with the observed value of $\De^2_\mR \sim 10^{-9}$
\cite{Komatsu:2010fb}, it is necessary for the critical density to be
of the order of $10^{-9} \rho_{\rm Pl}$.  This seems to contradict LQC
where the critical energy density is usually assumed to be of the order
of the Planck density, in which case the matter bounce scenario would
be ruled out in LQC.  However, it is important to remember that the
numerical value of the critical energy density must ultimately be
derived from loop quantum gravity, and this remains an open problem.
The critical energy density may turn out to be much smaller than
expected, in which case the matter bounce scenario would be viable
in LQC.

\subsection{The Equation of State $P = -\ep \rho$}
\label{ss.scal-eps}

A slight red tilt to the spectrum of scalar perturbations is obtained
by working with the equation of state $P = -\ep \rho$, where
$0 < \epsilon \ll 1$.  We will not go through all of the details of
the calculation here as they are a straightforward extension of what
is done earlier in this section.

It is a relatively simple task to determine the spectrum of scalar
perturbations for this new equation of state (the calculations
simplify considerably if terms of order $\ep^2$ and higher
are dropped).  The resulting spectrum has a scalar index of
\be 
n_s = 1 - 12 \ep,
\ee
which given the observed scalar index of $n_s = 0.968 \pm 0.012$
\cite{Komatsu:2010fb}, implies that we must have
\be \label{epsilon}
\ep \approx 0.003,
\ee
which validates the approximation $\ep \ll 1$.

Thus, it is possible to match the observed red tilt of the scalar perturbation
spectrum by choosing an appropriate matter field.  In the next section, where
we shall study tensor perturbations, we will exclusively work with a background
where the equation of state is $P = - \ep \rho$, as this is the relevant
setting whose predictions agree with observations.

\section{Tensor Perturbations}
\label{s.tens}

For tensor perturbations, we will follow the same procedure as in the
previous section, i.e., impose quantum vacuum initial conditions before
the bounce and use matching conditions in order to determine the
form of the tensor perturbations after the bounce.

One difference with the previous section is that we will work in
the setting where the equation of state is $P = -\ep \rho$, with
$0 < \ep \ll 1$.  This is the background that gives a slight red tilt
to the spectrum of scalar perturbations.  In addition, as we shall
see, for tensor perturbations the limit of $\ep = 0$ gives a qualitatively
different result so it is important to work from the start with a
nonzero value for $\ep$.  Because $\ep$ is very small, we will neglect
all terms that are of the order of $\ep^2$ or smaller.

To first order in $\ep$, the scale factor and the proper time
in the classical theory, in terms of the conformal time, are given
by
\be 
a(\eta) = \left( (1 - 6 \ep) \f{2 \pi G \rho_c}{3} \eta^2
\right)^{1 + 3 \ep},
\ee
\be 
t(\eta) = \left( 6 \pi G \rho_c \right)^{1 + 3 \ep}
\left( \f{1 - 2 \ep}{3} \eta \right)^{3 (1 + 2 \ep)}.
\ee
While it is possible to expand the exponents around $\ep = 0$, this
is not necessary and would merely complicate calculations later.

\subsection{The Contracting Branch and the Bounce}
\label{ss.tens-cont}

In the classical limit, the Mukhanov-Sasaki equation for tensor
perturbations with an equation of state $P = - \ep \rho$ is
given by
\be 
\mu'' + \left( k^2 - \f{2 (1 + 9 \ep)}{\eta^2} \right) \mu = 0,
\ee
and choosing the initial conditions to be the quantum vacuum state
\eqref{initial} gives
\be \label{mu-initial}
\mu(\eta) = \sqrt{ \f{-\pi \hbar \eta}{4} } H_n^{(1)} (-k \eta)
\ee
in the classical regime of the contracting branch of the cosmology
with
\be 
n = \f{3}{2}  + 6 \ep.
\ee

As before, we will link the solutions in the contracting and expanding
branches via the leading order solution in $k$ by using the integral
form of the modified Mukhanov-Sasaki equation,
\begin{align}
\mu(\eta) =& D_1 z_T + D_2 z_T \int^\eta \f{\dd \b\eta}{z_T^2}
- k^2 z_T \int^\eta \f{\dd \b\eta}{z_T^2}
\int^{\b\eta} \dd \b{\b\eta} \, z_T \, \mu \nn \\ &
+ \f{2 k^2}{\rho_c} z_T \int^\eta \f{\dd \b\eta}{z_T^2}
\int^{\b\eta} \dd \b{\b\eta} \, \rho \, z_T \, \mu,
\end{align}
and solving for the two leading order terms in $k$.

For $\om = - \ep$ in LQC,
\be 
z_T(t) = \f{\left[ 6 \pi G \rho_c (1 - 2 \ep) t^2 + 1
\right]^{\tf{5}{2} + \tf{3\ep}{2}}}
{\sqrt{6 \pi G \rho_c (1 - 2 \ep) t^2 - 1}},
\ee
and therefore we find that the two leading order terms in $k$
of $\mu$ are
\begin{align} \label{mu-formal}
\mu(t) = \,& D_1 z_T(t) - \Bigg[ \f{\ep}{\sqrt{6 \pi G \rho_c}}
\left( \arctan \sqrt{6 \pi G \rho_c} \, t + \f{\pi}{2} \right) \nn \\ &
\quad + \f{(1 - \ep) t} {\left[ 6 \pi G \rho_c (1 - 2 \ep) t^2
+ 1 \right]^{1 + \ep}} \Bigg] \, D_2 z_T(t),
\end{align}
where the integration constant is again chosen in order to simplify
the matching procedure.  Then, in the regime $t \ll -1 / \sqrt{6 \pi G \rho_c}$,
the expression for $\mu(t)$ simplifies to
\be 
\mu(t) = D_1 a(t) - \f{(1 - 2 \ep) D_2 a(t)}
{[6 \pi G \rho_c (1 - 2 \ep)]^{1 + \ep} t^{1 + 2 \ep}},
\ee
and therefore, for the two solutions to match, we must have
\begin{align} 
D_1 = \, & \sqrt \f{8 \hbar}{9} \left( \f{3}{8 \pi G \rho_c}
\right)^{1 + 3 \ep} k^n, \\
D_2 = \, & i \sqrt \f{9 \hbar}{32} \, \left( \f{8 \pi G \rho_c}{3}
\right)^{1 + 3 \ep} \, k^{-n},
\end{align}
where we have dropped the nonexponential dependence of $\ep$ in the prefactors
as this will not be relevant for our calculations.

\subsection{The Expanding Branch}
\label{ss.tens-exp}

In order to determine the spectrum of gravitational waves in the
classical regime of the expanding branch after the bounce, it is
necessary to match the classical solution
\be 
\mu(\eta) = \sqrt{\eta} \left[ E_1 H_n^{(1)} (k \eta)
+ E_2 H_n^{(2)} (k \eta) \right]
\ee
to leading order in $k$ with the leading order terms of the formal
solution \eqref{mu-formal} in the regime $t \gg 1/ \sqrt{6 \pi G \rho_c}$,
where the two leading order terms have the form
\begin{align} 
\mu(t) = & \left( D_1 - \sqrt \f{\pi}{6 G \rho_c} \, \ep D_2 \right) a(t) \nn \\ &
\quad - \f{(1 - 2 \ep) D_2 a(t)} {[6 \pi G \rho_c (1 - 2 \ep)]^{1 + \ep} t^{1 + 2 \ep}}.
\end{align}
From this expression, and the previous calculations, one immediately
sees that
\begin{align} \label{mu-final}
\mu(\eta) = & \sqrt{\pi \hbar \eta} \Bigg[
- i \f{\ep}{32 \, k^{2n}} \sqrt \f{27 \pi}{2 G \rho_c}
\left( \f{8 \pi G \rho_c}{3} \right)^{2 + 6 \ep} \!\!\!
J_n(k \eta) \nn \\ & \quad
+ \f{1}{2} H_n^{(2)} (k \eta) \Bigg],
\end{align}
where the prefactors are understood to be accurate to leading order in
$k$.  Again, just as in the case of scalar perturbations, a new mode is
created by the bounce.  As the amplitude of the tensor perturbations is
of the order of $\ep$, this will give a small tensor to scalar ratio.

From the expression of $\mu$, it is possible to determine $h$,
\be \label{h-final}
h = \f{\mu}{a} \sim 
-i \f{\ep \pi}{2} \sqrt \f{\hbar}{2}
\left( \f{8 \pi G \rho_c}{3} \right)^{\tf{1}{2} + 3 \ep}
k^{-\tf{3}{2} - 6 \ep},
\ee
where the first equality holds in the classical regime and we have kept
only the constant part of the leading order term in $k$ on the
righthand side.

The amplitude of the tensor perturbations is then given by%
\footnote{There is an extra factor of
$32 \pi G$ in the definition of $\Delta^2_h$
for dimensional reasons, and another additional
factor of $2$ to account for the two
polarizations.  See e.g.\ \cite{Baumann:2009ds}
for details.}
\begin{align} 
\De^2_h(k) &= 64 \pi G \f{k^3}{2 \pi^2} |h|^2 \nn \\ &
= \ep^2 \, \f{32 \pi^2 G^2 \hbar \rho_c}{3}
\left( \f{8 \pi G \rho_c}{3 k^2} \right)^{6 \ep},
\end{align}
and thus we get a spectrum of tensor perturbations with a small
amplitude and an almost scale-invariant spectrum with $n_T = - 12 \ep$.

Finally, it is possible to determine the tensor to scalar
ratio, given by
\be 
r = \f{\De^2_h}{\De^2_\mR} = 96 \, \ep^2 \approx 9 \times 10^{-4},
\ee
where we have inserted the value of $\ep$ in \eqref{epsilon}, which
was determined by the observed tilt of the spectrum of scalar
perturbations.  The predicted value for $r$ is so small that, if the
matter bounce scenario is the correct one, we do not expect to observe
a primordial gravitational wave background until the precision of
astronomical observations increases by an order of magnitude or two.

It is worth noting that it is important to work with the equation of
state $P = -\ep \rho$ for tensor perturbations as the predictions
are significantly different than for a pure dust matter field with
$P = 0$: for the case $P = 0$, no new mode appears in \eqref{mu-final}
and then the resulting spectrum would have a strong blue tilt of
$n_T = 6$ (and therefore primordial gravitational waves would not
be observable as their amplitude would be proportional to $\lp^6 k^6$,
an extremely small factor for all relevant modes).  Thus, the
small deviation from a dust fluid plays an important role as it
generates the new mode in \eqref{mu-final} which is scale-invariant
and the only one which is potentially observable today in this
scenario.

\section{Discussion}
\label{s.dis}

In loop quantum cosmology, quantum gravity effects modify both the Friedmann
equations and the Mukhanov-Sasaki equations, which govern the dynamics of
the homogeneous background and of the linear perturbations respectively.
These modifications become important at large curvature scales, and especially
at the bounce point.  Therefore, it is important to include these effects
in any setting where the cosmological bounce plays a role, and this is what
has been done here for the matter bounce scenario.

The matter bounce in loop quantum cosmology gives a scale-invariant spectrum
for scalar and tensor perturbations, and a slight red tilt is obtained for
the equation of state $P = - \ep \rho$, with $0 < \ep \ll 1$.  The same tilt
is predicted for the scalar and tensor modes.  As an identical tilt for the
scalar and tensor modes is not expected in inflationary models, this is
one way the two scenarios can be differentiated.

The other main results are that the observed amplitude for scalar perturbations
is obtained for a value of the critical energy density of
$\rho_c \sim 10^{-9} \rho_{\rm Pl}$, and a tensor to scalar ratio is
predicted to be $r \approx 9 \times 10^{-4}$.  Note that there are some
important differences between these predictions for the matter bounce
scenario and those given in \cite{Finelli:2001sr} where quantum gravity
effects were not included, particularly regarding the amplitudes of the
spectra and their relative importance.

Although the value of $\rho_c$ is usually assumed to be within one order of
magnitude of the Planck energy density, this is a quantity that should be
derived from loop quantum gravity.  Thus, until the relation between the
full theory of LQG and its cosmological sector is better understood, the critical
energy density remains an unknown, and may turn out to be considerably
smaller than expected.  In any case, it is clear that for the matter bounce
scenario to be viable in LQC we must have $\rho_c \sim 10^{-9} \rho_{\rm Pl}$.

One of the key predictions here is the small tensor to scalar ratio, which
is proportional to $\ep^2$ (and is therefore related to $n_s$).  The
modifications to the scalar and tensor Mukhanov-Sasaki equations are slightly
different, and this plays a major role in the matter bounce scenario as one
of its effects is the much smaller amplitude of the tensor perturbations
than that predicted from classical considerations.  The reason why the
amplitude of the tensor perturbations is so small (and vanishes in the
limit $\ep \to 0$) is that $z_T$, defined in \eqref{zt}, becomes imaginary
near the bounce.  Therefore, the integral $\int \dd \eta / z_T^2$ can vanish
as some portions of it will be negative.  On the other hand, the integrand
in $\int \dd \eta / z^2$ [see \eqref{scal-z}] is always positive so the
integral cannot vanish.  This is why the tensor to scalar ratio is suppressed
for the matter bounce scenario in LQC.  The difference between the LQC
modifications to the scalar and tensor Mukhanov-Sasaki equations is one
way that quantum gravity effects can significantly change results obtained
in a purely classical setting.

It is interesting to compare these results to those obtained for the matter
bounce scenario in different cosmological models.  Two matter bounce cosmologies
that give similar predictions are (i) with a Lee-Wick type scalar field that
causes a bounce \cite{Cai:2008qw} (the case where the Lee-Wick scalar is
non-minimally coupled to gravity is studied in \cite{Qiu:2010ch}), and (ii)
in the setting of the $f(T)$ generalization of teleparallel gravity
\cite{Cai:2011tc}.  It is possible to incorporate the matter bounce scenario
in both of these cases, and then the spectra for the scalar and tensor
perturbations after the bounce will be scale-invariant.  However, the main
shortcoming of both of these models is that they predict a tensor to scalar
ratio that is approximately unity (at least for the typical implementation
of these models), thus violating the bound of $r \lesssim 0.25$ given in
\cite{Komatsu:2010fb}.  Another setting the matter bounce scenario has been
considered in is that of Bohmian quantum cosmology \cite{Peter:2006hx}.
Although a small tensor to scalar ratio is obtained in this case, the spectra
for the scalar and tensor perturbations are expected to have a slight blue
(rather than red) tilt.  It would be possible to obtain a slight red tilt by
using a scalar field rather than dust as the matter content in the Bohmian
quantum cosmology, but then the resulting scalar to tensor ratio would be
close to unity.  In the matter bounce scenario, it is often difficult to
obtain the two characteristics of a red tilt in the spectrum of scalar
perturbations and a small tensor to scalar ratio.  As we have seen, this is
not the case in LQC.

A more promising matter bounce alternative is studied in \cite{Cai:2012va},
where a combination of ingredients coming from the matter bounce and ekpyrotic
cosmological scenarios is used in order to obtain scale-free perturbations and
also a small scalar to tensor ratio.  In this setting, the small value of $r$
arises because the sound speed of the scalar perturbations becomes negative
for a short time and this generates exponential growth in the amplitude of the
scalar perturbations.  Since the speed sound of the tensor perturbations
remains positive, their magnitude is unchanged and thus $r$ is small.
Note that this is the opposite mechanism of what occurs in LQC: in LQC, it
is the negative value of $z_T^2$ during the bounce which damps the amplitude
of the tensor modes, while in this other scenario it is the negative speed
sound that generates growth in the scalar modes.
An important point is that this cosmology can be differentiated from the one
studied in this paper by observations as the predicted scalar to tensor ratio
in \cite{Cai:2012va} is $10^{-5}$ (although the predicted value of $r$ in this
setting is model dependent to some extent), almost two orders of magnitude
smaller than what is predicted in LQC.  This is the main difference between
the predictions of the two cosmologies.

Finally, it is reasonable to expect that quantum gravity effects could
also give nontrivial corrections to the classical predictions of other
cosmological models, including inflation.  As it is known how to include
inflation in LQC \cite{Ashtekar:2011rm, FernandezMendez:2012vi}, it is
important now to study how quantum gravity effects could affect the
standard inflationary predictions.  Some work has already been done in
this direction, see for example \cite{Bojowald:2011iq, Agullo:2012sh}.

As the effects of LQC and other quantum cosmology models on the CMB and
primordial gravitational waves are better understood, and the observations
continue to improve, it will become possible to differentiate between (i)
alternative cosmological scenarios (inflation, matter bounce, etc.), and
(ii) the various quantum cosmology theories (LQC, string cosmology, etc.),
and thus determine which combination of the two is realized in nature.

\acknowledgments

The author would like to thank
Ivan Agull\'o,
Thomas Cailleteau
and
William Nelson
for helpful discussions.
This work was supported by Le Fonds qu\'eb\'ecois de la recherche
sur la nature et les technologies.



\end{document}